\documentclass[prb,twocolumn]{revtex4}

\usepackage{graphicx,amsmath}
\usepackage{natbib}

\begin{document}

\title{Spectral Properties of  Holstein and Breathing Polarons}
\author{
C.\ Slezak$^{1}$, A.\ Macridin$^{1}$, G.\ A.\ Sawatzky$^{2}$,
M. \ Jarrell$^{1}$ and
T.A.\ Maier$^{3}$}

\address{
$^{1}$University of Cincinnati, Cincinnati, Ohio, 45221, USA \\
$^{2}$University of British Columbia, 6224 Agricultural Road, Vancouver,  BC
V6T 1Z1,
Canada\\
$^{3}$Oak Ridge National Laboratory, Oak Ridge, Tennessee, 37831, USA }

\date{\today}

\begin{abstract}

We calculate the spectral properties of the one-dimensional Holstein
and breathing polarons using the self-consistent Born approximation.
The Holstein model electron-phonon coupling is momentum independent
while the breathing coupling increases monotonically with the phonon
momentum. We find that for a linear or tight binding electron
dispersion: i) for the same value of the dimensionless coupling the
quasiparticle renormalization at small momentum in the breathing
polaron is much smaller, ii) the quasiparticle renormalization at
small momentum in the breathing polaron increases  with  phonon
frequency unlike in the Holstein model where it decreases,  iii) in
the Holstein model the quasiparticle dispersion  displays a kink and a
small gap at an excitation energy equal to the phonon frequency
$\omega_0$ while in the breathing model it displays two gaps, one at
excitation energy $\omega_0$  and another one at  $2\omega_0$.  These
differences have two reasons: first, the momentum of the relevant
scattered phonons increases with increasing  polaron momentum and
second, the breathing bare coupling is an increasing function of the
phonon momentum. These  result in an effective electron-phonon
coupling for the breathing model which is an increasing function of
the total polaron momentum, such that the small momentum polaron is in
the weak coupling regime  while the large momentum one is in the
strong coupling regime.  However the first  reason does not hold if
the free electron dispersion has low energy  states separated by large
momentum, as in a higher dimensional system for example, in which
situation the  difference between the two models becomes less
significant.

\end{abstract}
\maketitle

\section{Introduction}

Interest in the spectral properties of strongly coupled
electron-phonon systems has increased due to the discovery of a kink
in the quasiparticle
dispersion~\cite{cukrev,lanzara,bogdanov,kaminski} measured by
angle-resolved  photoemission (ARPES) experiments in cuprate
superconductors.  The energy of this kink is characteristic of a
specific optical phonon mode.  Furthermore, a peculiar isotope-effect,
manifesting  in a shift of the quasiparticle dispersion at high energy
upon oxygen isotope substitution, was  recently measured in the high
T$_c$ materials~\cite{gweon}, undoubtedly showing that the
electron-phonon interaction plays an important role in the spectral
properties of cuprates.

The electron-phonon interaction has been investigated in a variety of
models~\cite{holstein,su,froehlich}. Although the ground state
properties of these models received great attention, less was paid to
the spectral features.  Noticeable exceptions include the
Holstein~\cite{engelsberg,mishchenko} and the
Fr\"{o}lich~\cite{prokofev} models which are characterized by a
momentum independent or a weakly momentum dependent electron-phonon
coupling.  However, in many systems, such as cuprates superconductors
or organic materials, the electron-phonon coupling is strongly
momentum dependent.  This can give rise to distinctive properties of
the single particle spectral features, as we will discuss in this
paper.  In particular we will consider a sinusoidal dependence of the
coupling on the phonon momentum, which is realized in many systems
including the SSH coupling in polyacetilens~\cite{su} and the coupling
to the breathing mode in  cuprate
superconductors~\cite{abanov,friedl,kee}.

The treatment of a multi-electron system coupled with phonons is an
extremely complex and difficult problem.  In general, phonons mediate
an effective attraction between electrons and consequently the system
becomes susceptible to various kinds of instabilities~\cite{bcs,jahn},
which might have a significant effect on the photoemission spectra.
However, in this paper we investigate the effect of phonons on a
single quasiparticle only, thus  neglecting  electron-electron or
electron-hole scattering processes at the Fermi surface. We believe
this to be suitable in describing the properties at energy or
temperature scales larger than those associated with the ground state
instability, e.g.\ the normal state of superconductors. 

In fact, as different investigations have
shown~\cite{mishchenko,shen}, the polaron models, which consist of a
single electron interacting with phonons~\cite{landau}, capture much
of the physics seen in the ARPES experiments on materials with
significant electron-phonon interaction~\cite{lanzara,hengsberger}.
This includes the kink in the quasiparticle dispersion observed at the
phonon characteristic frequency.  One expects that the single particle
description of the influence of electron-phonon coupling would  be
valid to describe the photoelectron or inverse photoelectron spectral
function in insulators and semiconductors in which the bands are
either full or empty.  However things are more complicated for a
strongly correlated insulator in which the conductivity gap is a
result of electron correlations and the material is an insulator in
spite of having a half filled band. In systems described by the Hubbard
or t-J models the single hole spectral function will be  influenced by
the interaction with spin fluctuations.  This results in strongly
dressed quasiparticles even without the electron-phonon coupling, as
calculations employing self consistent Born approximation
(SCBA)~\cite{shushkov} and exact diagrammatic quantum Monte Carlo
(QMC)~\cite{mishchenko1}  shows.  The later calculation also shows
that the SCBA, which neglects all the magnon crossing diagrams, is a
good approximation to the quasiparticle mass renormalization and its
dispersion. When the  phonons are considered, in the simplest
approximation we can neglect all the crossing diagrams involving
phonons and magnons and treat the interaction with phonons in a single
particle fashion by taking the hole band dispersion as that given by
the t-J model calculations. The approximations involving the polaron
models under investigation here are done in this spirit.

In this paper we investigate the spectral properties of two  polaron
models: The one-dimensional Holstein (H) model~\cite{holstein} and a
one-dimensional version of the breathing model (B)~\footnote{The
breathing mode actually can be properly defined only at the zone
corner in 2D and at $k=\pi$ in 1D.  However in this paper we extend
this nomenclature for all the $k$ points, even if this is not entirely
correct.}, relevant for cuprates
superconductors~\cite{roesch,pintschovius}.  The electron-phonon
coupling in the H Hamiltonian is a constant, thus independent of the
phonon momentum.  The B model has a sinusoidal momentum dependent
electron-phonon coupling, being small (large) for  scattering with
small (large) momentum phonons.  In both cases we consider
dispersionless optical phonons with  frequency  $\omega_0$.  Although
for the sake of simplicity our calculations are done in one dimension,
the conclusions and the qualitative properties of the spectra are
independent of dimensionality.

The difference between the properties of the two models emphasizes the
importance  of  momentum dependent couplings.  The specific momentum
dependence of the B bare electron-phonon coupling results in  an
effective  coupling which is an increasing function of the total
polaron momentum, such that the small momentum polaron is in the weak
coupling regime  while the large momentum one is in the strong
coupling regime. This might be germane to the peculiar behavior of the
high energy quasiparticle dispersion~\cite{gweon} or the  temperature
dependency of the  photoemission linewidth~\cite{pothuizen} in cuprate
superconductors.

The method we use in our calculations is the self-consistent Born
approximation (SCBA)\cite{martinez}.  Although this method  is an
uncontrolled approximation which  neglects electron-phonon vertex
corrections in the self-energy calculation, the results at small
coupling are in good agreement with numerically exact quantum Monte
Carlo calculations.  Because the contribution of configurations with
an infinite number of phonons is considered, this approximation
manages to capture  the physics at large momentum, such as the gap and
the flattening of the quasiparticle dispersion at the phonon
energy~\cite{larsen}. It can  easily  provide information about
excited states and thus determine the polaron spectral properties,
unlike most QMC methods~\cite{prokofev,raedt,kornilovitch} fit  to
calculate only the ground state properties.

The paper is organized as follows. In  Sec.~\ref{model} we introduce
the H and B Hamiltonians. The dimensionless electron-phonon coupling
is defined in Sec.~\ref{epcoupl}. The SCBA method is discussed in
Sec.~\ref{methode}. The results are presented in Sec.~\ref{results}
and their significance is discussed in Sec.~\ref{discussion}. A short
summary and the conclusions are given in Sec.~\ref{conclusion}.

\section{Formalism}

\subsection{Model}
\label{model}

The Holstein Hamiltonian in real space is given by
  
\begin{equation}
\label{eq:Hham}
H_H=-t\sum_{<ij>}(c^\dagger_{i}c_{j}+h.c.)+\omega_0 \sum_{i} b^\dagger_i b_i
+g \sum_{i}n_i x_i~,
\end{equation}

\noindent where $c_{i}$ ($b_{i}$) is the electron (phonon)
annihilation operator at site $i$. The first term describes the
electron kinetic energy.  The second term describes a set of
independent oscillators with frequency $\omega_0$ at every site.  The
electron-phonon coupling in the Holstein model is local and is
described by the last term of Eq.~\ref{eq:Hham} where the electron
density $n_i=c^\dagger_{i}c_{i}$ couples with the lattice displacement
$x_i=\frac{1}{\sqrt{2 M \omega_0}}(b^\dagger_i+b_i)$ with strength
$g$.

Our breathing Hamiltonian  is a one-dimensional version of the model
which describes the coupling of Zhang-Rice (ZR) singlets~\cite{zhang}
with the Cu-O bond-stretching vibrations in high T$_c$
superconductors~\cite{pintschovius,roesch}.  In cuprates a ZR singlet
is a bound state between a hole on the Cu and a hole on the four
neighboring O atoms. It's energy is stabilized by the Cu-O
hybridization term and therefore is influenced by the Cu-O distance.
In our one dimensional model we consider a set of independent,
in-between sites oscillators (the analogue of the O atoms)  which
modulate the charge carrier's (the analogue of the ZR singlet) on-site
energy.  Therefore we define the B Hamiltonian as 
\begin{eqnarray}
	\label{eq:Bham} 
	H_B = &
	-t\sum_{<ij>}(c^\dagger_{i}c_{j}+h.c.)+ \omega_0 \sum_{i}
	b^\dagger_{i+\frac{1}{2}} b_{i+\frac{1}{2}} \\ \nonumber + & g
	\sum_{i}n_i(x_{i-\frac{1}{2}}-x_{i+\frac{1}{2}})~~.
\end{eqnarray}

\noindent Both the H and  the B model can be written in the momentum
representation as

\begin{eqnarray}
\label{eq:kham}
H _{(H,B)}=  \sum_{k} \epsilon(k) c^{\dagger}_k c_k  + 
\sum_q \omega_0  b^{\dagger}_q b_q + \\ \nonumber
\frac{1}{\sqrt{N}} 
\sum_{k,q} \gamma_{(h,b)} (q) c^{\dagger}_{k-q} c_k (b^{\dagger}_q+b_{-q})
\end{eqnarray}

\noindent where 
\begin{equation}
\label{eq:hgamma}
\gamma_H(q)=\frac{g}{\sqrt{2 M \omega_0}}
\end{equation}

\noindent is the H electron-phonon coupling and
\begin{equation}
\label{eq:bgamma}
\gamma_B(q)=-i \frac{2g}{\sqrt{2 M \omega_0}} \sin{\frac{q}{2}}
\end{equation}

\noindent is the B electron-phonon coupling.  Notice that in the
momentum representation the H coupling is a constant and the B
coupling is an increasing function of the phonon momentum for small
momenta.

In Eq.~\ref{eq:Hham} and Eq.~\ref{eq:Bham} the free electron part of
the Hamiltonian was introduced as a tight-binding hopping. However in
order to study the influence of electron dispersion on the polaron
properties in this paper we also employ  calculations which consider
different forms of electron dispersion.

\subsection{Electron-phonon coupling}
\label{epcoupl}

In this paper, we  define the dimensionless 
electron-phonon coupling as
the ratio of the lattice deformation energy of a localized electron
and the kinetic (delocalization) energy of the electron. 
This definition is encountered mostly in polaron studies focusing on the transition 
from large to small polarons~\cite{kornilovitch,ku}.
For the Holstein model the deformation energy is given by (see Eq.~\ref{eq:eph})
\begin{equation}
\label{eq:Eph}
{E_p}_H=  \frac{1}{2M\omega_0}~ \frac{g^2}{\omega_0}~,
\end{equation} 
\noindent and  the dimensionless coupling is defined as
\begin{equation}
\label{eq:lambdah}
\lambda_H =  \frac{2{E_p}_H}{W}=\frac{1}{2M\omega_0}~ \frac{g^2}{tz \omega_0 }
\end{equation}
\noindent  where the half-bandwidth $W/2$ is taken as a measure of the free electron
kinetic energy. For a simple tight binding dispersion $W=2zt$, where 
$z$ is the coordination number.

Analogously, for the breathing model the lattice deformation energy is (see
Appendix~\ref{appbreathing})
\begin{equation}
\label{eq:Eb}
{E_p}_B=  \frac{1}{2M\omega_0}~ \frac{z g^2}{\omega_0}~,
\end{equation} 
\noindent and the dimensionless coupling is hence defined as
\begin{equation}
\label{eq:lambdab}
\lambda_B =  \frac{2{E_p}_B}{W}= \frac{1}{2M\omega_0}~ \frac{z g^2}{tz \omega_0 }~.
\end{equation}
\noindent  Compared to the Holstein case, the lattice deformation energy for
the breathing model has an extra factor of $z$ as can be seen in Eq.~\ref{eq:Eb},
which results form the fact that the electron interacts with $z$
neighboring oscillators.

There are several other definitions for the  dimensionless electron-phonon coupling 
throughout the literature. For instance in the BCS theory the definition of the dimensionless coupling
is $\lambda=V N(0)$, where $V$  and  $N(0)$ are the effective electron-electron attraction 
and respectively the density of states  at the Fermi level.
In the weak coupling regime the effective interaction at 
small frequency  or in the antiadiabatic limit ($\omega_0/t \rightarrow \infty$) is 
\begin{equation}
\label{eq:BCS}
V(q,\omega \ll \omega_0)=   \frac{2 |\gamma(q)|^2}{\omega_0 }~.
\end{equation}
\noindent If the integration over all $q$ momenta is considered for the breathing model, then
\begin{eqnarray}
\label{eq:lambdah1}
\lambda_H^{BCS}=&\frac{2 N(0)}{2M\omega_0 } \frac{g^2}{\omega_0}\\
\label{eq:lambdab1}
\lambda_B^{BCS}=&\frac{2 N(0)}{\omega_0 } \frac{1}{2 \pi } \int dq |\gamma(q)|^2= 
\frac{2 N(0)}{2M\omega_0}~ \frac{2 g^2}{ \omega_0 }~.
\end{eqnarray} 
\noindent When choosing the density of states  $N(0)=1/W$,
this definition of the dimensionless coupling coincides with our definition.

In the Migdal-Eliashberg theory of superconductivity, 
$\lambda^{ME} =2 \alpha^2F(\omega_0)/\omega_0$ where only scattering processes at the Fermi surface are 
considered~\cite{allen,scalapino} in $\alpha^2F(\omega)$. This definition is equivalent 
to\cite{scalapino}
\begin{equation}
\label{eq:lambda2}
\lambda^{ME}=\frac{m*}{m}-1
\end{equation} 
\noindent where $m*$ is the quasiparticle renormalized effective mass at the Fermi surface. 
$\lambda^{ME}$ can be directly determined from experiments 
since  $\alpha^2F(\omega)$ and $m*/m$ can be measured in tunneling~\cite{vedeneev}, 
neutron~\cite{shapiro} or respectively in the ARPES experiments~\cite{lanzara}. 
In our case, the effective mass at the bottom of the band (which is the
zero energy state of our system and in many respects similar to the Fermi surface in a many-electron system)
can be determined  from the equation
\begin{equation}
\label{eq:mass1}
(E(k)-E(0))(1-\frac{\partial \Sigma}{\partial \omega}(0,E(0))=
\epsilon(k)-\epsilon(0)+\frac{\partial \Sigma}{\partial k}(0,E(0))k~
\end{equation}
\noindent valid at small $k$. The ratio of the effective mass and the un-renormalized mass is
\begin{equation}
\label{eq:mass2}
\frac{m}{m*}=\frac{E(k)-E(0)}{\epsilon(k)-\epsilon(0)}=\frac{1}{1-\frac{\partial
\Sigma}{\partial \omega}(0,E(0))}=Z_0~.
\end{equation}
\noindent Therefore, we have 
\begin{equation}
\label{eq:lambdame}
\lambda^{ME}=\frac{1}{Z_0}-1 \approx - \frac{\partial \Sigma}{\partial \omega}(0,E(0))~.
\end{equation}
This definition directly relates the quasiparticle weight $Z_0$ to the  coupling constant $\lambda^{ME}$.
In the   first order  perturbation theory one gets
\begin{equation}
\label{eq:lambdapt}
\lambda^{ME}=\frac{1}{2\pi}\int dq \frac{|\gamma(q)|^2}{(\epsilon(0)-\omega_0 -\epsilon (q))^2}~.
\end{equation} 
\noindent By inspecting Eq.~\ref{eq:lambdapt} one can see that  the most important contribution 
to the integral comes form small $q$ where the denominator is small. For the B model this will 
introduce a significant difference between $\lambda^{ME}_B$ and our $\lambda_B$ 
(Eq.~\ref{eq:lambdab}). Unlike our definition which assumes average over all possible phonon 
momenta, $\lambda^{ME}_B$ is determined by small $q$ phonon scatterings which  are characterized 
by small $\gamma(q)$ in the B model.

The definition of $\lambda^{ME}$ used in the Migdal-Eliashberg theory
was intended to describe the properties at the Fermi level or at small energy excitation.
It does not properly describe the physics at larger energy since the most
relevant  scattering processes in this case imply larger phonon momenta as we will
discuss in the next two sections.

\subsection{Method}
\label{methode}

In the SCBA the electron self-energy is obtained by summing over all non-crossing 
diagrams. In this approximation
the  calculation of the self-energy can be reduced
to the following set of equations:

\begin{figure}
\centerline{\includegraphics*[width=3.3in]{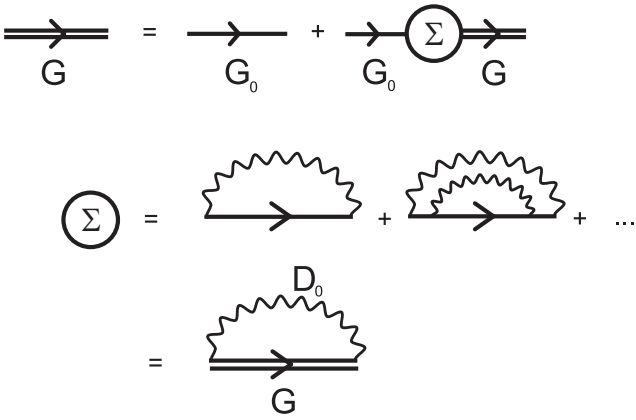}}
\caption{Schematic representation of the Dyson equation ({\em upper}).
In  the SCBA approximation  the self-energy is a summation of all non-crossing diagrams ({\em middle})
and can be determined self-consistently using Eq.~\ref{eq:sigma} ({\em bottom}).}
\label{fig:selfdig}
\end{figure}

\begin{equation}
\label{eq:Dyson}
G(k,\omega)= \frac{1}{\omega-\epsilon(k)-\Sigma(k,\omega)}~,
\end{equation}

\begin{equation}
D(q,\Omega)= \frac{2\omega_0}{\Omega^2-\omega_0^2}~,
\end{equation}

\noindent and
\begin{equation}
\label{eq:sigma}
\Sigma(k,\omega)= \frac{1}{(2 \pi)^2}\int {\rm d}\Omega \int {\rm d}q ~|\gamma(q)|^2 D(q,\Omega)
G(k-q,\omega-\Omega)~, 
\end{equation}
\noindent where $G(k,\omega)$ and $D(q,\Omega)$ are the free electron and respectively the free 
phonon Green's functions (see also Fig.~\ref{fig:selfdig}). 
\noindent The frequency integration in Eq.~\ref{eq:sigma} can be 
 explicitly completed  and one gets
\begin{equation}
\label{eq:selfcons}
\Sigma(k,\omega)=  \frac{1}{2 \pi}\int {\rm d}q\frac{|\gamma(q)|^2}{\omega-\omega_0-
\epsilon(k-q)-\Sigma(k-q,\omega-\omega_0)}
\end{equation}
\noindent where $\omega_0$ is the phonon frequency. 
Eq.~\ref{eq:selfcons} is solved self-consistently and the 
spectral function is determined as $A(k,\omega)= - \frac{1}{\pi} \text{Im} G(k,\omega)$.
The overlap of the polaron state $| \nu_k>$ (the lowest energy state at a particular momentum $k$ which
in general is an isolated pole in the Green's function) with the free electron state, 
$c^{\dagger}_{k}|0\rangle$
\begin{equation}
\label{eq:zdef}
Z_k=|\langle \nu_k|c^{\dagger}_{k}|0\rangle |^2~,
\end{equation} 
\noindent is called the quasiparticle weight and can be calculated as~\cite{martinez}
\begin{equation}
\label{eq:zcalc}
Z_k=\left. \frac{1}{1-\frac{\partial \Sigma(k,\omega)}{\partial \omega}} \right|_{\omega=E_k}~.
\end{equation}

\begin{figure}
\centerline{\includegraphics*[width=3.3in]{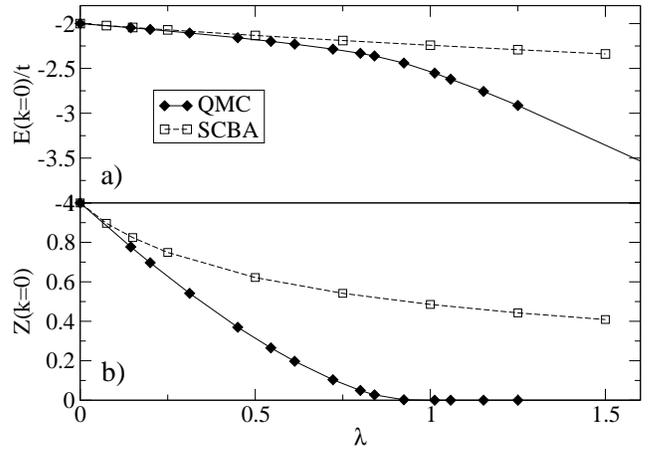}}
\caption{Numerically exact QMC (diamonds) and SCBA  (squares) results for polaron
energy $E_0$ (a) 
and quasiparticle weight $Z_0$ (b) versus the dimensionless electron-phonon coupling $\lambda$ 
(see Eq.~\ref{eq:lambdah}) at $k=0$. Holstein model with $t=1$ and
$\omega_0=0.1~t$ and $2M=1$.}
\label{fig:QSlam}
\end{figure}

\begin{figure}
\centerline{\includegraphics*[width=3.3in]{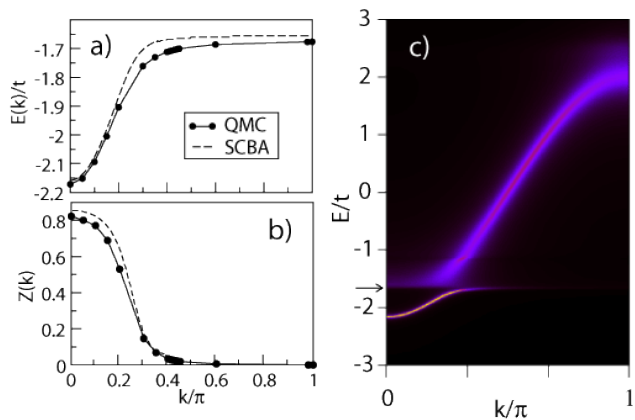}}
\caption{ Holstein polaron dispersion $E(k)$ (a) and  quasiparticle weight $Z(k)$ (b)
calculated with SCBA (dotted line) and QMC (circles).  
c) Spectral representation obtained with SCBA. 
Everywhere $\lambda_H=0.25$ and $\omega_0=0.5~t$.}
\label{fig:qmcscba}
\end{figure}

The neglect of vertex corrections in the SCBA
results in a failure of this approximation in the intermediate and strong coupling regime.
This is apparent in Fig.~\ref{fig:QSlam} where a comparison of  SCBA and exact QMC
results for the H model energy and quasiparticle weight at $k=0$ is shown.
In fact the polaron properties at $k=0$ are better approximated in a
regular Rayleigh-Schrodinger (RS) perturbation  theory than in SCBA~\cite{mahan}. However,
at large $k$ where the  polaron energy approaches the phonon frequency, 
the RS perturbation theory fails and the SCBA  which
considers configurations involving a large number of phonons 
provides good results. In order to show that, we compare 
the SCBA results with the exact ones obtained by using
the diagrammatic quantum Monte Carlo technique\cite{macridin,thesmac} in Fig.~\ref{fig:qmcscba}. 
As can be seen in Fig.~\ref{fig:qmcscba} -a, the dispersion becomes  flat
when the  polaron energy gets close to the the phonon frequency
$E_0+\omega_0$, $E_0$ being the bottom of the polaronic band.
The flattening is accompanied by a strong reduction of the quasiparticle weight,
as shown in  Fig.~\ref{fig:qmcscba} -b.
The SCBA slightly underestimates the energy corrections in the flat 
dispersion region but otherwise captures all these features successfully.

The H polaron physics  at small coupling
has been investigated with different techniques
for many years and is rather well understood now. The polaron 
character changes from almost free electron at
the bottom of the band to a {\em one phonon plus one electron}
state in the flat dispersion region where almost all the momentum is carried by the phonon. 
The states below the one phonon threshold $E_0+\omega_0$ are bound states\cite{prokofev}, 
characterized by delta peaks in the spectral representation. Above it, there is a  continuum 
of states and the electron self-energy acquires a finite imaginary part. 
However, at large energies relative to $\omega_0$ the self-energy's imaginary part is small and the spectral 
representation is characterized by sharp peaks with a dispersion close to 
the free electron one. In a photoemission  experiment  the dispersion 
of the high intensity peak would exhibit something  resembling 
a kink, although a small gap  at the phonon energy 
and sudden onset of broadening appears at the same time.
The spectral function at momenta corresponding to energies
well below the phonon frequency  will consist of a sharp peak followed
by a broad satellite, such as it is observed in surface studies of $Be$~\cite{hengsberger} for instance.  
This can be seen in   Fig.~\ref{fig:qmcscba} c) where 
a false color plot of spectral intensity obtained with SCBA is shown (see also Fig.~\ref{fig:spectb} -a).
The physics discussed in this paragraph (i.e.\ the dispersion flattening and the  gap in the polaron
spectrum)  is presumably true even for very small electron-phonon couplings. 
However, for such small couplings  these
features are too small to be captured experimentally or even
numerically.

\section{Results}
\label{results}

\begin{figure}
\centerline{\includegraphics*[width=3.3in]{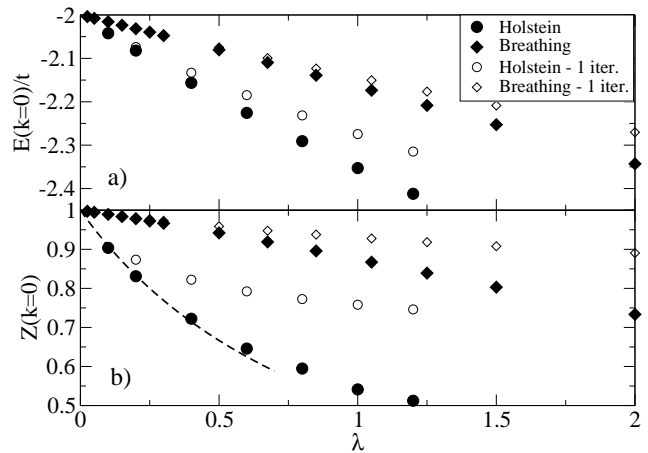}}
\caption{H (circles) and B (diamonds) polaron energy (a) and 
quasiparticle weight (b)  versus $\lambda$ at $k=0$ and $\omega_0=0.2 t$. The open symbols are
the results obtained after only one iteration (first order in $D$). The dashed line in (b) corresponds to
$\lambda^{ME}=\frac{1}{Z_0}-1$ (see Eq.~\ref{eq:lambdame}).}
\label{fig:ZvsLam}
\end{figure}

In our calculations we take $t=1$ and  $2M=1$. For the one-dimensional case 
the coordination number $z=2$.

In Fig.~\ref{fig:ZvsLam} -a and -b we show the polaron energy $E_0$ and respectively the
quasiparticle weight $Z_0$ versus the dimensionless coupling constant $\lambda$ at $k=0$. For the 
H model, the decrease of $Z_0$ with increasing $\lambda$ is not very different from the one 
given by the ME theory (dashed line in Fig.~\ref{fig:ZvsLam}-b given by Eq.~\ref{eq:lambdame}), 
showing that the ME definition of $\lambda$ is similar to ours. However, for the B model 
the quasiparticle weight $Z_0$ and the polaron energy $E_0$ decrease much slower with increasing  
$\lambda$. As mentioned in the previous section this is due to the small momentum scattering 
processes implied in the renormalization of $Z_0$ (see Eq.~\ref{eq:lambdapt}) and, similarly, in the 
determination of the self-energy. It is worth pointing out that, unlike the H polaron, 
even for values of $\lambda \approx 1$, the B polaron remains in the weak coupling regime 
and hence the difference between the fully convergent SCBA (full symbols) and the first order perturbation 
theory (i.e.\ only the first SCBA  iteration, empty symbols) is small.    
Another interesting feature is that for the same value of $Z_0$ the B polaron energy is 
lower than the H one, showing that the ratio between the energy renormalization and the quasiparticle weight 
renormalization is different for the two models.

\begin{figure}
\centerline{\includegraphics*[width=3.3in]{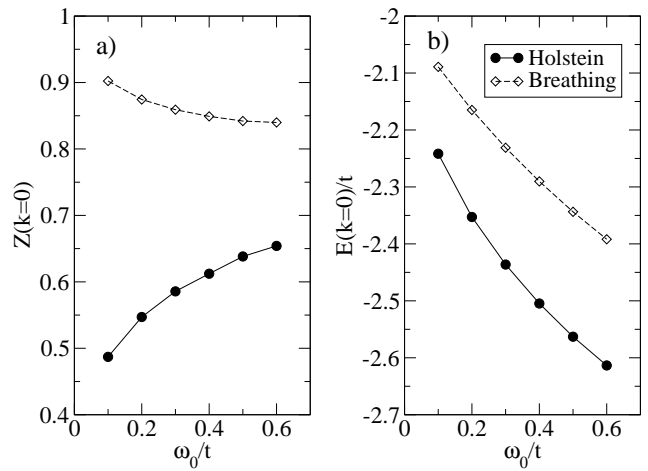}}
\caption{The  quasiparticle
weight $Z_0$ (a) and the polaron energy $E_0$ (b) at zero momentum versus phonon frequency
$\omega_0$ for the H and B  models at $\lambda_H =0.2$ and respectively
$\lambda_B =1.3$. }
\label{fig:omega}
\end{figure}

\begin{figure}
\centerline{\includegraphics*[width=3.3in]{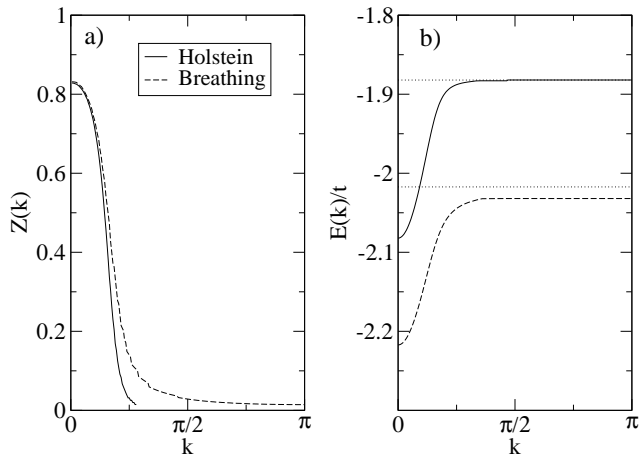}}
\caption{Quasiparticle weight $Z(k)$ (a) and the polaron energy $E(k)$ (b) for the 
H and B model for $\omega_0=0.2~t$ at coupling $\lambda_H =0.2$ and $\lambda_B
=1.3$ respectively. The dotted horizontal lines in (b) marks the first phonon
threshold energy $E_0+\omega_0$.} 
\label{fig:selfe}
\end{figure}

Another important difference between the two models is the
dependence of the polaron properties at  the band bottom on the phonon frequency  $\omega_0$. 
While for the H case an increase of  $\omega_0$ results in an increase of the 
quasiparticle weight $Z_0$ the opposite behavior is seen for the B model. This is illustrated
in Fig.~\ref{fig:omega}-a. The reason for the reduction of $Z_0$ with increasing $\omega_0$ in the 
B model can be easily understood by noticing (see Eq.~\ref{eq:lambdapt}) that a larger value of $\omega_0$ 
reduces the  importance of the $q$ dependence in the polaron properties  calculation. 
As discussed earlier, the strong 
momentum dependent coupling is responsible for the weak $Z_0$ renormalization of the B
polaron and thus an increase of $\omega_0$ would result in a larger effective coupling and 
implicitly in a smaller $Z_0$.

\begin{figure}
\centerline{\includegraphics*[width=3.3in]{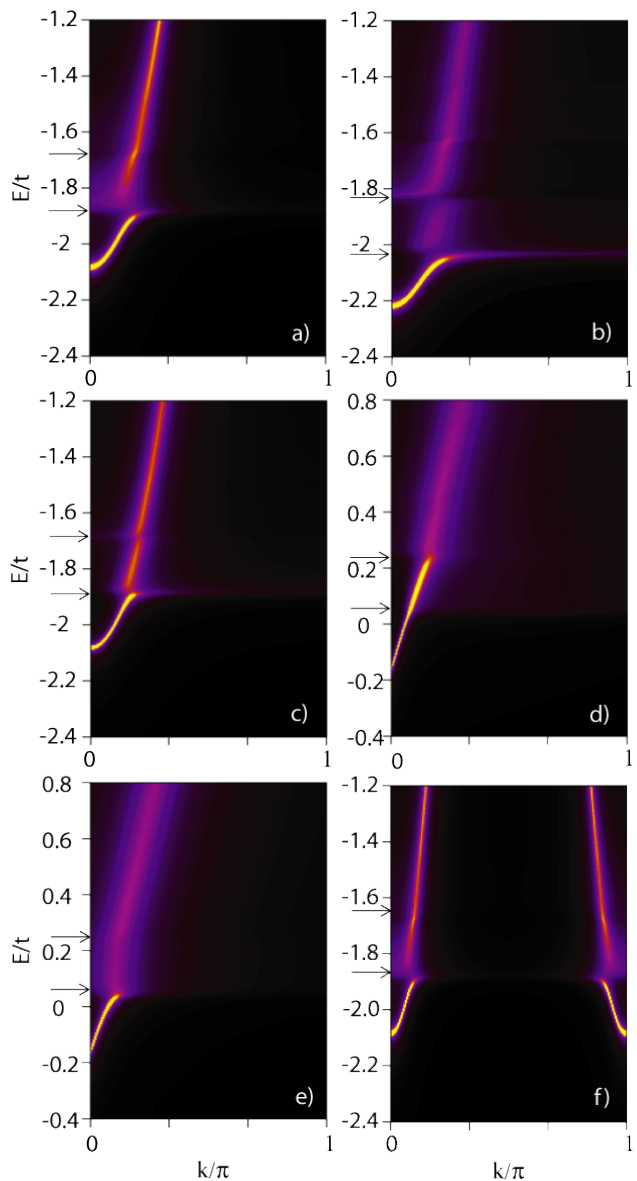}}
\caption{ Spectral representation for:
a) H model, $\lambda_H=0.2$, tight binding dispersion,
b) B model, $\lambda_B=1.3$, tight binding dispersion,
c) B model, $\lambda_B=0.5$, tight binding  dispersion,
d) B model, $\lambda^{lin}_B=1.6$, linear dispersion with $v_F=t$ (Eq.~\ref{eq:lambdab_lin}),
e) H model, $\lambda^{lin}_H=0.9$, linear dispersion with $v_F=t$,
f) H and B model, $\lambda_H=\lambda_B=0.2$, with dispersion $\epsilon(k)=-2t \cos(2k)$.
Everywhere $\omega_0=0.2t$.
The arrows indicate the first ({\it lower}) and the second
({\it upper}) phonon threshold energies corresponding to $E_0+\omega_0$ and  respectively
$E_0+2\omega_0$.} 
\label{fig:1dkink}
\end{figure}

\begin{figure}
\centerline{\includegraphics*[width=3.3in]{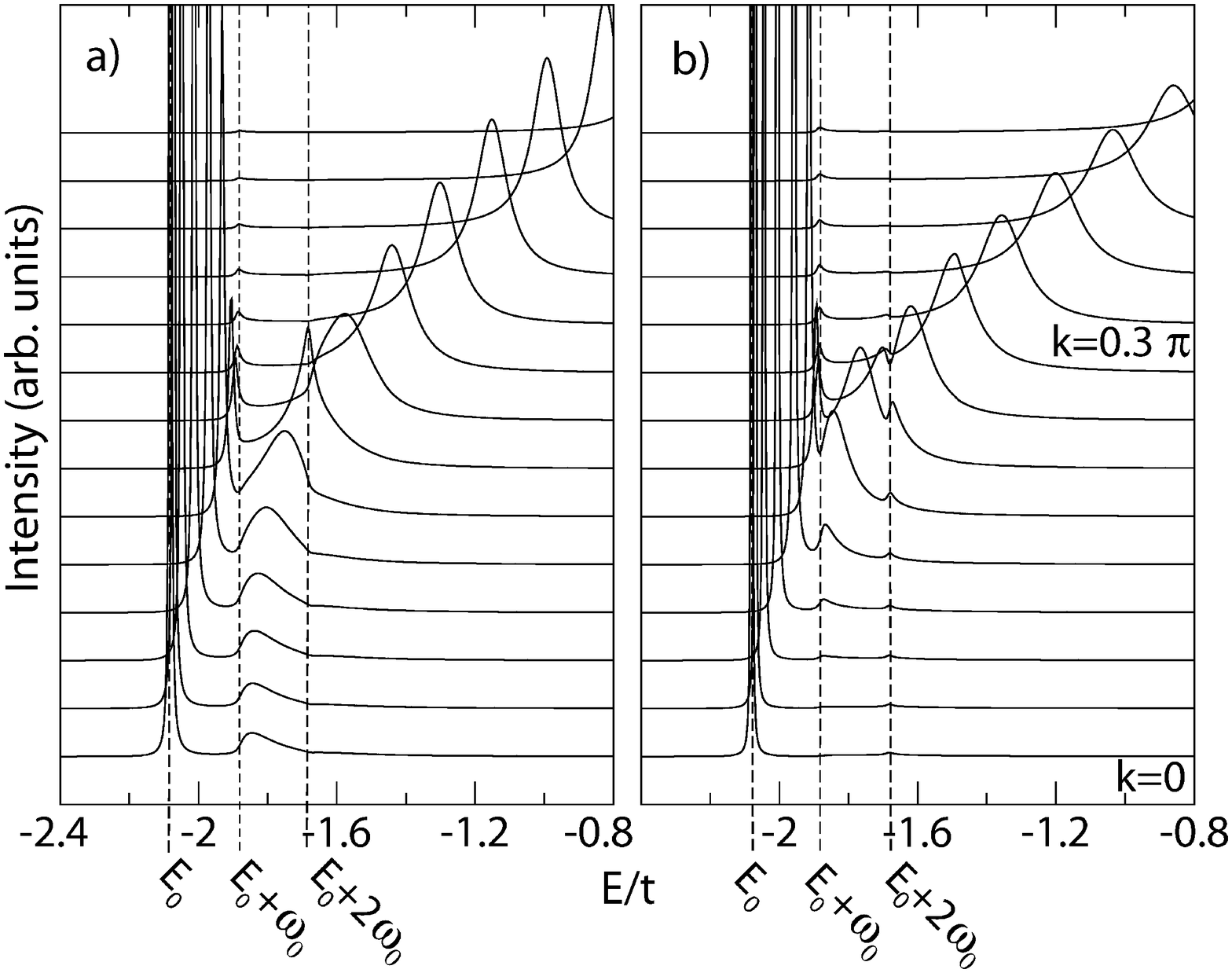}}
\caption{Spectral function (energy distribution curves)  for a) H model with $\lambda_H=0.2$ and
b) B model with $\lambda_B=0.5$. Free electron tight binding dispersion and  $\omega_0=0.2t$\
is considered.}
\label{fig:spectb}
\end{figure}

\begin{figure}
\centerline{\includegraphics*[width=3.3in]{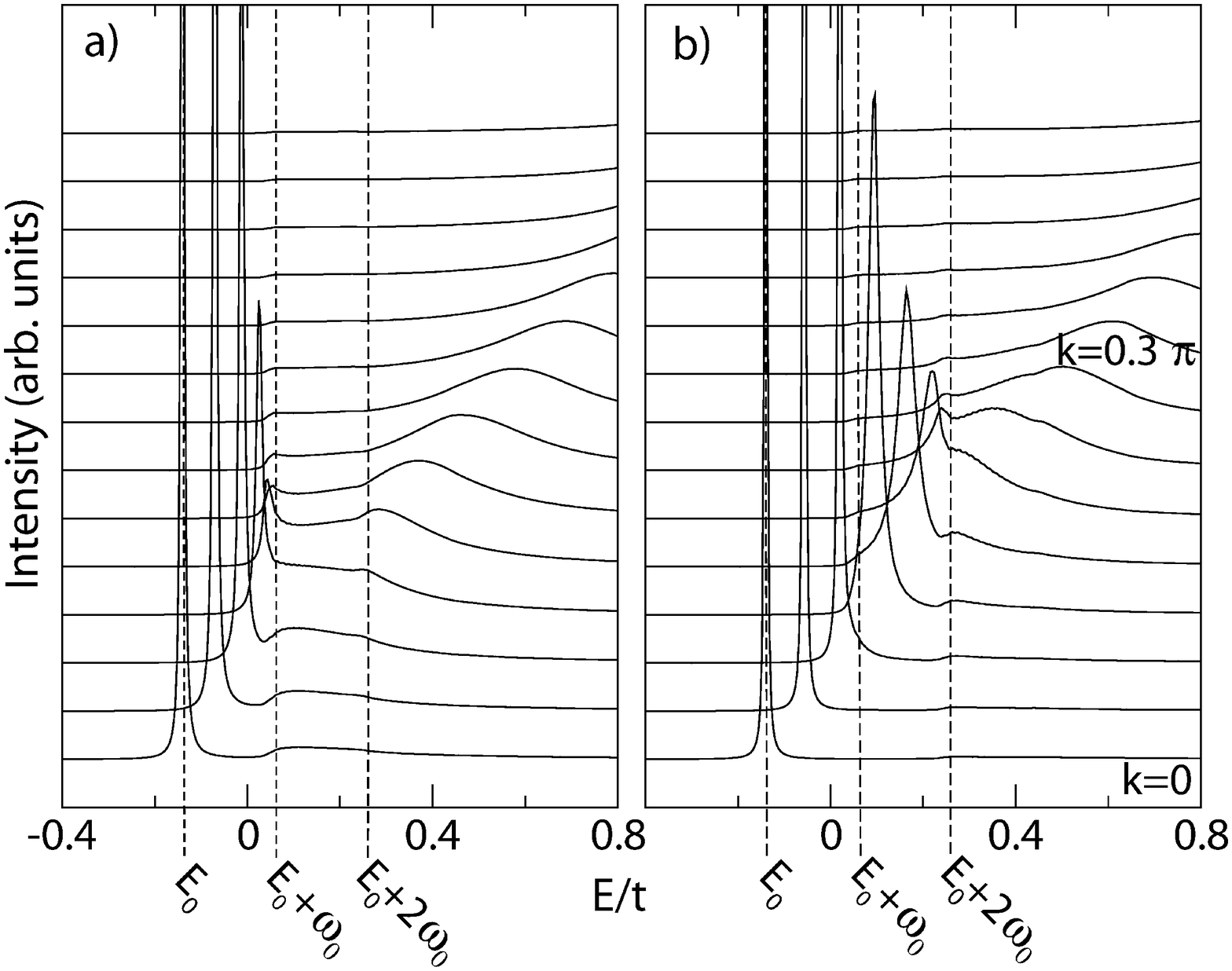}}
\caption{Spectral function (energy distribution curves) for a) H model with $\lambda_H^{lin}=0.9$ and
b) B model with $\lambda_B^{lin}=1.6$.  Free electron linear dispersion 
($v_F=t$) and  $\omega_0=0.2t$ is considered.}
\label{fig:speclin}
\end{figure}

Aside from the different $\lambda$ and $\omega_0$ dependency  of the
two models  at zero momentum, the momentum dependent properties  also
exhibit different behaviors. This is shown in
Figs.~\ref{fig:selfe},~\ref{fig:1dkink},~\ref{fig:spectb} and
~\ref{fig:speclin} where the $k$ dependent properties for the two
models are illustrated.

In Fig.~\ref{fig:selfe} and Fig.~\ref{fig:1dkink}-a and -b we have chosen the value of 
$\lambda$  such that both models yield the same quasiparticle weight at the bottom of the band.
Thus, the choice of $\lambda_H=0.2$ and $\lambda_B=1.3$ results in
$Z_{0h}=Z_{0b}=0.83$, implying that 
both models are in the weak coupling regime. As mentioned earlier, the B polaron 
energy at $k=0$ is lower. At large $k$, just below the first phonon threshold energy 
$E_0+\omega_0$, both polarons display a flat dispersion and a reduced quasiparticle weight.
However, the B polaron quasiparticle weight at large $k$ is substantially larger (for 
instance $Z_B(k=\pi) \approx 0.014$ within a numerical precision of $10^{-3}$ and 
$Z_H(k=\pi)<10^{-3}$ ), making the B polaron state at large $k$ distinguishable in the 
spectral  plot (Fig.~\ref{fig:1dkink} -b) in contrast to the H one 
(Fig.~\ref{fig:1dkink}-a). At energies larger than $E_0+\omega_0$ the spectral intensity of the 
B quasiparticle is much smaller than the H model one, unlike 
the situation at the band bottom where both models have the same $Z_0$.
This large momentum behavior points to a stronger effective coupling for the B model at 
large $k$. 

While the dispersion of both models displays a gap at $E_0+\omega_0$, the B polaron shows 
a second gap at the second phonons threshold energy $E_0+2 \omega_0$.
This can be seen in Fig.~\ref{fig:1dkink}-b but also occurs for smaller values of the 
dimensionless coupling as shown in Fig.~\ref{fig:1dkink}-c for the value of
$\lambda_B=0.5$.
This value was chosen such that the ground state energy $E_0$ of the B model is equal to that of 
the Holstein polaron one shown in Fig.~\ref{fig:1dkink}-a.
The situation can be even clearer visualized by comparing  Fig.~\ref{fig:spectb}-a
with  Fig.~\ref{fig:spectb}-b, where the energy distribution curves (EDC)
for H and respectively B cases are shown.

An even more interesting effect is noticed if a linear dispersion
for the free electron is considered
 
\begin{equation}
\label{eq:elin}
\epsilon(k)= v_F |k|~,
\end{equation}

\noindent with a value of $v_F$ close to one or larger. In this case one can take $v_F$ to be a 
measure of the free electron kinetic energy~\footnote{We set the lattice constant equal to 
unity.} 
and thus define the dimensionless coupling as 

\begin{equation}
\label{eq:lambdab_lin}
\lambda^{lin}_B =  \frac{{E_p}_B}{v_F}=\frac{1}{2M\omega_0}~ \frac{2 g^2}{ v_F \omega_0 }~.
\end{equation}

\noindent The resulting B polaron dispersion is shown in Fig.~\ref{fig:1dkink}-d (see also
the corresponding EDC plot in Fig.~\ref{fig:speclin}-b). While 
it displays a gap at $E_0+2 \omega_0$, no distinguishable gap or kink can be seen at the first 
phonon threshold energy $E_0+\omega_0$. 
This free-electron like behavior of the polaron at $E_0+\omega_0$ is due to the fact that the 
physics there is  determined by very small $q$ scatterings, characterized by small 
coupling strength $\gamma(q)$, originating from the rapid increase of the electron energy with $k$.
At larger energy, close to $E_0+2 \omega_0$, the relevant phonon momenta $q$
implied in the scattering are larger and the physics is consequently determined
by a larger effective coupling.  As a result a noticeable kink appears at this energy in the 
spectrum. 
This effect (i.e.\ kink at $E_0+2 \omega_0$ but no noticeable one at  $E_0+\omega_0$)  
is a result of an electron-phonon coupling which is an increasing function of the polaron momentum
and it is hence not seen in the Holstein model even for the case of a linear electronic dispersion
(see Fig.~\ref{fig:1dkink}-e and the corresponding EDC plot in Fig.~\ref{fig:speclin}-a).

The differences between the two models discussed above are a
consequence of two effects: i) the strong $q$ dependence of the bare
electron-phonon coupling $\gamma(q)$ in the B model and ii) the
polaron properties at small $k$ are  most strongly influenced by the
small momentum $q$ phonons.  However the second statement is not true
if the free electron dispersion has low energy states separated by
large $q$ as we will discuss in the next section
(Sec.\ref{discussion}).  In order to show this we choose a free
electron dispersion 

\begin{equation}
\label{eq:2kdisp}
\epsilon(k)=-2t\cos(2k)~,
\end{equation}

\noindent which is double degenerate with the lowest energy values at
$k=0$ and $k=\pi$ respectively.  We find that for this electronic
dispersion the differences between the H and the B model are very
small, less than $0.1\%$, and therfore not discernible in the
spectral representation plot shown in Fig.~\ref{fig:1dkink}-f.

The general features of the polaron spectral function illustrated in
Figs.~\ref{fig:spectb} and  \ref{fig:speclin} show a remarkable
resemblance with the  photoemission data in materials with significant
electron-phonon interactions~\cite{lanzara,hengsberger}. For momenta
which correspond to energies below the phonon frequency one can see a
sharp peak followed by a broad satellite.  At large momenta, the broad
satellite found at energies well above the phonon frequency sharpens
and follows the free electron dispersion while the intensity of the
sharp peak below $\omega_0$ vanishes. We want to point that this
picture, with the spectral function   determined by two main branches,
is quite different form the one incorrectly presented in some popular
textbooks~\cite{aschroft} where the spectrum is characterized by a
single quasiparticle peak which changes its dispersion slope  when
crossing the phonon energy threshold.

\section{Discussion}
\label{discussion}

For a tight-binding or a linear dispersion, the B polaron properties at small momentum
are determined by small $q$ scatterings characterized by small $\gamma(q)$. 
Let's consider the first order self-energy diagram contribution

\begin{equation}
\label{eq:self1}
\Sigma(k,\omega)=\frac{1}{2\pi}\int dq \frac{|\gamma(q)|^2}{ \omega-\omega_0 -\epsilon (k-q)}~.
\end{equation} 

\begin{figure}
\centerline{\includegraphics*[width=2in,height=2in]{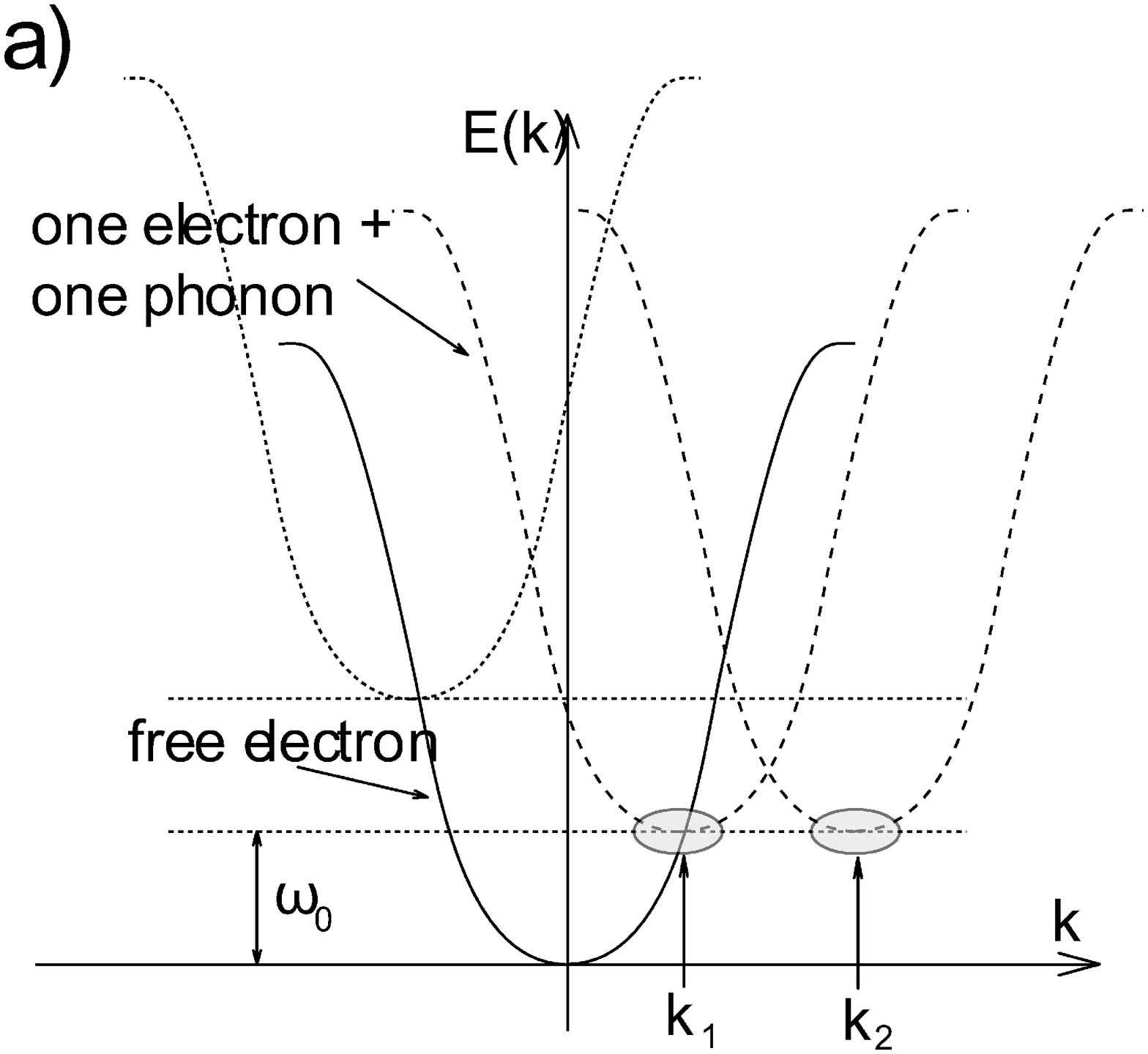}
\includegraphics*[width=1.3in,height=1.5in]{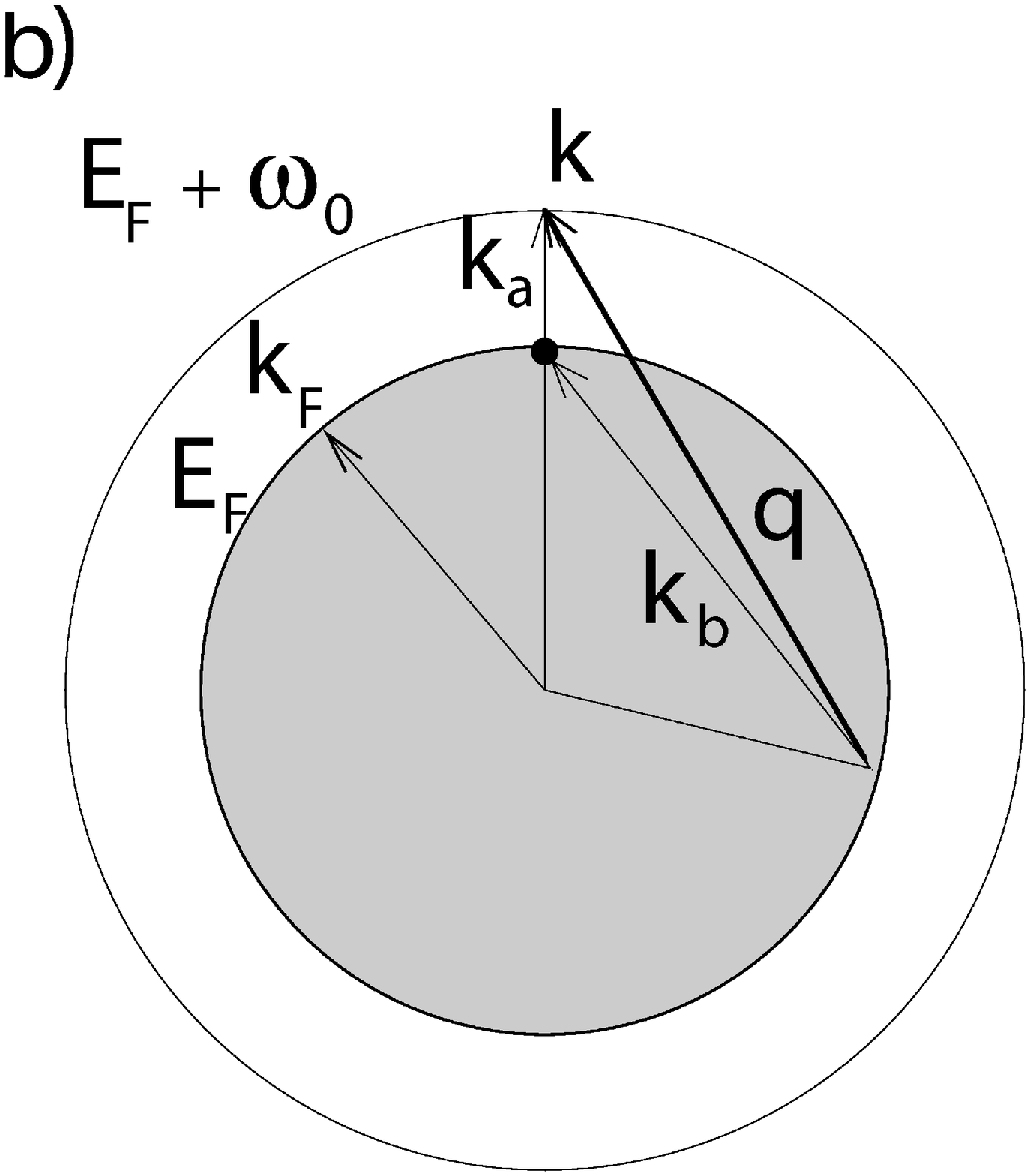}}
\caption{a) Energy levels for zero coupling ($g=0$). The solid line shows the non-interacting 
electron dispersion, the dashed ones are {\em one phonon + one electron} states and the dotted one 
{\em two  phonons + one electron}, etc. $k_1$ is the momentum where the electron dispersion reaches 
$\omega_0$ and has the most  significant contribution to the scattering at the first phonon threshold. $k_2$ is the value of 
momentum with significant contribution to the scattering at the second phonon threshold.
b) For the polaron with momentum $k$ the scattering with  momentum $q$  has a significant 
contribution. 
Although the polaron momentum is close to $k_F$, thus $k_a=k-k_F$ is  small, 
 $q=k_a+k_b$ is large because $k_b$ which connects two points on the Fermi
surface can be large.}
\label{fig:diag} 
\end{figure}

\noindent The most significant contribution below the one phonon threshold,
i.e.\ when
$\omega < \epsilon(0)+ \omega_0$, is given by the values of transferred momentum $q$ for which
$\epsilon(k-q)$ is minimum which implies $q \approx k$. The value of $k$ where the polaron energy 
reaches the one-phonon threshold is small  ($k_1$ in Fig.~\ref{fig:diag}-a) and thus,  
the relevant scatterings which determine the physics below $E_0+ \omega_0$ occur at 
small momentum $q$, implying a small effective coupling. 

For a linear dispersion with $v_F \approx 1$ or larger, the value of $k$ where the first phonon 
threshold energy is reached is even smaller. Besides, the electronic density of states at small 
momentum, which is proportional to the number of relevant scatterings, is
also much smaller then for a 1D tight-binding dispersion~\footnote{DOS for the
linear dispersion is constant $= 1/v_F$ while for a 1D tight-binding
dispersion it is infinite at $k=0$ .}. These two conditions yield a very small effective coupling
at the first phonon threshold. Consequently,  even for large values of $\lambda$ as defined in 
Eq.~\ref{eq:lambdab_lin}, the polaron dispersion exhibits an extremely narrow gap, hardly  
discernible in our calculation.

The B model properties at larger momentum indicate an increase of the effective electron-phonon
coupling with increasing the polaron momentum, due to the increase of the momentum  of the relevant 
scattered phonons.
For example, the most significant scattering 
at the two phonon threshold ($\omega \approx E_0+2 \omega_0$) are those for which 
$\epsilon(k-q) \approx \omega_0$. If $k_1$ is the value of the electron momentum for which
$\epsilon(k_1)= \epsilon(-k_1)=\omega_0$ the resulting values for the phonon momentum $q$ 
relevant in the scattering are  $q \approx k \pm k_1$. The larger $q$
solution, i.e.\  
$q \approx k +  k_1$ ($k_2$ in Fig.~\ref{fig:diag}-a), implies a larger effective coupling.

A similar analysis of the higher order self-energy diagrams leads to the same conclusion: The 
effective coupling in the B polaron model is increasing with momentum, and hence the small 
energy and momentum  B polaron is characterized by a small effective coupling while the high energy 
and large momentum properties are determined by a large effective coupling.
 Moreover, at low energy the contribution 
of the crossing diagrams is comparable to the contribution of the crossing diagrams in a H
model at small coupling. Thus, the SCBA solution is a good approximation for the low energy, small 
$k$, polaron. Nevertheless, the large momentum properties are characterized by large effective 
couplings and hence the SCBA approximation is questionable in that region.

The conclusions based on the analysis of  self-energy diagrams discussed above
implicitly assume that a small $\epsilon(k-q)$ implies small $q$, when the polaron momentum $k$ 
is small.
However, as we have shown in the previous section, if the
electron dispersion has low energy states separated by large
$q$  the above assumption is invalid. Thus the effective coupling at small energy
is substantially larger and besides there is no significant increase of the effective
coupling with polaron momentum or energy.

For example the strong dependence of the effective coupling on the momentum does not 
hold in a metallic system with a large Fermi surface. 
To see this, one can think of the Fermi surface 
as a degenerate ground state for the quasiparticle~\footnote{The physics at  energies
close to  $\omega_0$ relative to Fermi surface is presumably not influenced much by the electron
hole creation at the Fermi surface.}.
Unlike  the  non-degenerate ground state case where the small energy scattering is
restricted to small momentum, the relevant scatterings in the presence of a Fermi surface are
restricted to small momentum {\em plus} a vector which connects two points on the Fermi surface 
($q=k_a+k_b$ in Fig.~\ref{fig:diag} b)). Therefore scatterings with large momentum, for which 
$\gamma(q)$ is large, are relevant. This makes  the effective coupling to be large 
even at small energy. Because now at both low energy and high energy there
are relevant large momentum  scatterings  the dependence of the effective coupling strength on the 
energy  and momentum will  be modest.

Although dimensionality  plays an important role in determining the quantitative polaron 
properties~\cite{kornilovitch,ku}, the main features of the B polaron are a consequence of  
strong momentum dependence of the bare electron-phonon coupling. Therefore the 
main conclusions of this study  remain valid for 2D or 3D systems as long as the electron-phonon 
coupling has a similar $q$ dependence (increases with $q$).

The particularities of the B polaron discussed in this paper might be relevant for the 
weakly doped cuprates characterized by very small Fermi pockets around the 
($ \pm \pi/2, \pm \pi/2$) points in the Brillouin Zone which makes the situation very similar
to the one captured by the polaron model. The quasiparticles in this system have a four-fold 
degenerate ground state with the states separated by $(0,\pi)$ or $(\pi,0)$
vectors~\cite{shushkov}. The relevant 
phonon mode is believed to be the half-breathing one for which the electron-phonon coupling has a 
strong $q$ dependence, being small at small $q$ and large at  $(0,\pi)$. If for some reasons the 
scattering between the ($ \pm \pi/2, \pm \pi/2$) points is restricted or if its importance
is small because of an additional  $k$ dependence in the bare electron-phonon coupling $\gamma(k,q)$
the formed polarons will 
be characterized by a momentum and  energy dependent effective coupling. 
Consequently, this implies  interesting energy and temperature dependent properties 
such as a strongly temperature dependent quasiparticle photoemission 
linewidth~\cite{pothuizen}.
However, if the scattering between the ($ \pm \pi/2, \pm \pi/2$) states is relevant, the 
$q$ dependence of the coupling will not be very important 
and presumably a Holstein like coupling can be 
used to describe the physics as well, analogous to the 1D case discussed in 
Fig.~\ref{fig:1dkink}-f.

\section{Conclusions}
\label{conclusion}

In this paper we study the spectral properties of the one-dimensional Holstein and breathing 
polaron models. While the H model has a momentum independent electron-phonon 
coupling, the B coupling is a monotonic increasing 
function of the phonon momentum.
We find the renormalization of the quasiparticle properties at small momentum in the B 
model weaker than that in the H one for a dimensionless coupling constant defined as 
the ratio of the lattice deformation energy and the kinetic energy of the electron.
With increasing phonon 
frequency  $\omega_0$, the B  quasiparticle weight decreases in contrast to the H case where it increases.
The quasiparticle dispersion in the H model displays a kink and a small gap
at an excitation energy equal to $\omega_0$. On the other hand,
the B model dispersion,  besides the gap at $\omega_0$, 
exhibits another, more pronounced one at energy $2\omega_0$.
This is due to the following facts:  the momentum $q$ of the  relevant phonons 
in the scattering process increases with increasing the total polaron momentum 
and that the B coupling is an increasing function of $q$. As a consequence
the  renormalization of the B polaron increases with increasing  polaron momentum. 
However the first fact is  dependent on the free electron dispersion and does not hold 
if it has low energy  states separated by large momentum. 
In that case the difference between the B and the H polaron becomes less significant.

The specific momentum dependence of the B bare electron-phonon 
coupling results in  an effective  coupling 
which is an increasing function of the total polaron momentum, such
that the small momentum  polaron properties are weak coupling while the 
large momentum ones are strong coupling. This might be relevant for
explaining the peculiar behavior  of the high energy quasiparticle
dispersion~\cite{gweon} or the  temperature dependency of the  
photoemission linewidth~\cite{pothuizen} in
cuprate superconductors.

\section*{Acknowledgments} The work was supported by NSF grants DMR-0312680 
and DMR-0113574, and by CMSN grant DOE DE-FG02-04ER46129.  We  acknowledge
Ohio Supercomputer Center where part of the computation was performed.  
G.~A.~Sawatzky acknowledges the financial support from the Canadian funding
organizations NSERC, CIAR and CFI.  Thomas Maier acknowledges support from the
Center for Nanophase Materials Sciences, Oak Ridge National Laboratory, which
is funded by the Division of Scientific User Facilities, U.S. Department of
Energy.

\appendix
\label{append}
\section{Effective coupling}

\subsection{Holstein}

The last two terms of the Holstein Hamiltonian (Eq.~\ref{eq:Hham}) can be diagonalized via the 
Lang-Firsov canonical transformation\cite{lang} defined by a unitary operator $e^S$ where

\begin{equation}
S=-\frac{g}{\omega_0 \sqrt{2 M \omega_0}}\sum_{i} n_i(b^{\dagger}_i
-b_i)~~.
\end{equation}

\noindent Using the expansion

\begin{equation}
\label{eq:expansion}
\tilde{A}=e^SAe^{-S}=A+[S,A]+\frac{1}{2}[S,[S,A]]+...
\end{equation}

\noindent we find for the transformed phonon annihilation operator:

\begin{equation}
\tilde{b}_i=b_i+\frac {g}{\omega_0\sqrt{2 M \omega_0}}n_i
\end{equation}

\noindent and similarly for the electron operator
\begin{equation}
\tilde{c}_i=c_i
e^{\frac{g}{\omega_0 \sqrt{2 M \omega_0}}(b^\dagger_i-b_i)}~~.
\end{equation}

\noindent The tilde mark is used to label the transformed operators.
The Holstein Hamiltonian in the new basis  can be written as

\begin{equation}
H=H_0+H_t
\end{equation}

\begin{equation}
\label{eq:Hnoth}
H_0=\sum_{i}\omega_0 \tilde{b}^\dagger_i  \tilde{b}_i -
\frac{g^2}{2 M\omega_0^2}\sum_{i}\tilde{n}_i^2
\end{equation}

\begin{equation}
H_t=-t\sum_{<ij>}(\tilde{c}^\dagger_i \tilde{c}_j X^\dagger _i X_j+H.c.)
\end{equation}

\noindent where

\begin{equation}
X_i=e^{-\frac{g}{\omega_0\sqrt{2 M \omega_0}}(\tilde{b}^\dagger_i-\tilde{b}_i)}~~.
\end{equation}

For a single electron the density term in Eq.~\ref{eq:Hnoth} simplifies, yielding

\begin{equation}
\label{eq:Hsingle}
H_0=\sum_{i}\omega_0 \tilde{b}^\dagger_i  \tilde{b}_i -
\frac{g^2}{2 M \omega_0^2}\sum_{i}\tilde{n}_i
\end{equation}

One can define a dimensionless coupling constant for the electron-phonon 
interaction as the ratio between the gained lattice deformation energy (see second term in $H_0$)

\begin{equation}
\label{eq:eph}
{E_p}_H=  \frac{1}{2M\omega_0}~ \frac{g^2}{\omega_0}~,
\end{equation} 

\noindent and the bare electron kinetic energy, taken to be the half-bandwidth,
 $W/2=zt$,

\begin{equation}
\label{lamholst}
\lambda_H= \frac{2{E_p}_H}{W}=\frac{1}{2M\omega_0}\frac{g^2}{z\omega_0t}~.
\end{equation}

\subsection{Breathing}
\label{appbreathing}

The electron-phonon interaction term in the 
B Hamiltonian (see last term in Eq.~\ref{eq:Bham}) can,
by changing the order of summation, be re-write it in the form

\begin{equation}
H_{int}=g \sum_{i} x_{i-\frac{1}{2}}(n_i-n_{i-1})~,
\end{equation}

\noindent where $x_i$ is the lattice displacement.
The last two parts of the Hamiltonian are once again
diagonalized via the unitary operator $e^S$ where

\begin{equation}
S=-\frac{g}{\omega_0 \sqrt{2 M \omega_0}}\sum_{i} n_i(b^{\dagger}_{i-\frac{1}{2}}
-b_{i-\frac{1}{2}})~~.
\end{equation}

\noindent Using the expansion of Eq.~\ref{eq:expansion} we find for the transformed phonon 
and electron annihilation operators:

\begin{equation}
\tilde{b}_{i-\frac{1}{2}}=b_{i-\frac{1}{2}}+\frac {g}{\omega_0\sqrt{2 M \omega_0}}
(n_i-n_{i-1})
\end{equation}

\begin{equation}
\tilde{c}_i=c_i
e^{\frac{g}{\omega_0 \sqrt{2 M \omega_0}}(b^\dagger_{i-\frac{1}{2}}-b_{i-\frac{1}{2}})}~~.
\end{equation}

\noindent Substituting these transformed operators we find in the
new basis

\begin{equation}
\begin{split}
\tilde{x}_{i-\frac{1}{2}}&=\frac{1}{\sqrt{2M\omega_0}}(\tilde{b}^{\dagger}_{i-\frac{1}{2}}+\tilde{b}_{i-\frac{1}{2}})\\
&=\frac{1}{\sqrt{2M\omega_0}}(b^{\dagger}_{i-\frac{1}{2}}+b_{i-\frac{1}{2}})
+\frac{2g}{2 M\omega_0^2}(n_i-n_{i-1})
\end{split}
\end{equation}

\noindent and for the breathing-Hamiltonian

\begin{equation}
\label{eq:Hnot}
H_0=\sum_{i}\omega_0 \tilde{b}^\dagger_i  \tilde{b}_i -
\frac{g^2}{2 M\omega_0^2}\sum_{i}(\tilde{n}_i-\tilde{n}_{i-1})^2
\end{equation}

\begin{equation}
H_t=-t\sum_{<ij>}(\tilde{c}^\dagger_i \tilde{c}_j X^\dagger _i X_j+H.c.)
\end{equation}

\noindent where

\begin{equation}
X_i=e^{-\frac{g}{\omega_0\sqrt{2 M \omega_0}}(\tilde{b}^\dagger_{i-\frac{1}{2}}-\tilde{b}_{i-\frac{1}{2}})}~~.
\end{equation}

In the case of a system only containing a single electron the density terms
contained in Eq.~\ref{eq:Hnot} simplify, 
yielding

\begin{equation}
\label{eq:HsingleB}
H_0=\sum_{i}\omega_0 \tilde{b}^\dagger_i  \tilde{b}_i -
\frac{2 g^2}{2 M \omega_0^2}\sum_{i}\tilde{n}_i~~.
\end{equation}

Hence, the lattice deformation energy is found to be

\begin{equation}
\label{eq:epb}
{E_p}_B=  \frac{1}{2M\omega_0}~ \frac{2 g^2}{\omega_0}~,
\end{equation} 

\noindent In contrast to the H model (Eq.~\ref{eq:eph}) we incur an additional factor of $2$
due to a coupling with two neighboring oscillators. For higher
dimensionality the number of neighboring oscillators is given by the coordination
number $z$, thus replacing the factor $2$ in Eq.~\ref{eq:epb} with $z$.

We thus define the dimensionless coupling constant for the electron-breathing-phonon 
interaction 

\begin{equation}
\lambda_B= \frac{2{E_p}_B}{W}=\frac{1}{2M\omega_0}\frac{zg^2}{z\omega_0t}~.
\end{equation}

\noindent This definition is identical to that found for the Holstein polaron besides the overall 
factor of $z$.


\end{document}